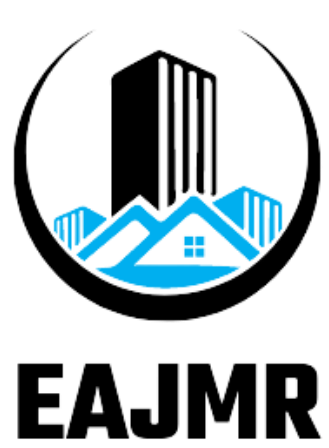

# A Path Model to Infer Mathematics Performance: The Interrelated Impact of Motivation, Attitude, Learning Style and Teaching Strategies Variables

**Marvin G. Pizon[1*], Shiryl T. Ytoc[2]**
[1]Applied Mathematics, College of Arts and Sciences, Agusan del Sur State College of Agriculture and Technology
[2]English Language, College of Arts and Sciences, Agusan del Sur State College of Agriculture and Technology

**ABSTRACT:** The present study aims at exploring predictors influencing mathematics performance. In particular, the study focuses in four subject's components such as motivation, attitude towards mathematics, learning style, and teaching strategies. A sample of 240 students from Agusan del Sur State College of Agriculture and Technology (ASSCAT) were involved in the study. Path analysis was used to test the direct and indirect relations between the predictors and mathematics performance. Based on the result, the calculation of reproduced correlation through path decompositions and subsequent comparison to the empirical correlation indicated that the path model fits the empirical data. Results also revealed that a large proportion of mathematics performance can be directly predicted from attitude towards mathematics, learning style, and teaching strategies. Moreover, attitude towards mathematics, learning style, and teaching strategies influence mathematics performance in direct and indirect ways.

## INTRODUCTION

Mathematics improvement is at the core of educational strategy in all over the world, but in Philippines still learners face with the difficulties in their math problem. In order to function in a mathematically literate way in the future, students must have a strong foundation in mathematics.

As stressed by Vintere and Zeidmane (2014), a high mathematics performance required in order to perform successfully professional task. However, the consequences of poor mathematics performance are serious for daily functioning and for profession development. The study of Alcuizar (2016) revealed that poor living condition and distance of the school affects low academic achievement and it was concluded by the study of Lacour and Tisington (2011) that poverty directly affects academic achievement. In spite of such continuing problem, many empirical studies are carried out to explore factors that affect the poor performance in mathematics. Although there are substantial research which as investigated the influences of motivation (Amrai et al, 2011; Guay et al, 2010), attitude towards mathematics (Farooq and Shah, 2008; Hemmings et al, 2011), learning style (Awang et al, 2017;Yildirim et al, 2008) and teaching strategies (Tulbure, 2012; Sariçoban and Saricaoğlu, 2008), on the mathematical performance. In general, these studies have looked at the separate effects of these components. Hence, we consider its impact on mathematics performance when motivation, attitude towards mathematics, learning style, and teaching strategies taken together and how much each can predict the mathematical performance in an integrated model.

The study of Damavandi et al., (2011) and Ganyaupfu (2013), the implication of different student learning style and teaching strategy has significant effect on the student learning achievement in mathematics  On the other hand, Mata, *et al*., (2012) in their study that students can develop their mathematics competence through their attitude towards mathematics, since they learns to associate positive experiences. Moreover, self-efficacy boost the students' mathematics achievement through mathematics motivation that improved mathematics performance of the students (Liu and Koirala, 2009).

Some factors affecting the mathematics performance had been studied in the past. However, researchers of this study aimed to create a model that would best predict to enhance the mathematics performance of the students and to establish the specific cause-and-effect among the motivation, attitude towards mathematics, learning style, and teaching strategies on their mathematics performance, causal modeling specifically the path analysis will be sufficient enough to address this concern because it sole purpose is provide estimates of the magnitude and significance of hypothesized causal connections between sets of variables.

## THEORETICAL REVIEW

The exogenous variables of this research were Motivation ($X_1$), Attitude towards Mathematics ($X_2$), Learning Style ($X_3$), and Teaching Strategies ($X_4$). Meanwhile, the Mathematics Performnce (Y) was identified as the endogenous variable. This study had looked into the potentials of using these variables based on the following literature reviews:

In the study of Adnan *et al*., (2013) and Vaishnav (2013) there exist a relationship between learning style and mathematics achievement. Also, the study of Damrongpanit(2014), revealed that students' learning styles and teachers' teaching styles had direct effect on mathematics performance.Additionally, the study of Shi (2011), revealed that learning styles correlated with learning strategies.

According to Skinner's (1945) Learning Theory, achievements vary among individuals due to several reasons: level of performance and aspirations of students depend on factors linked to the level of education of parents, family income and marital status of parents. The theory further emphasizes the importance of motivation, involvement in learning. Also, Coleman (2009) in his study revealed that students motivated and build attitude toward the subject, they would be more likely to do effort and therefore achieved higher scores.

The Social Learning Theory as cited by Joyce *et al* (2003) that explains the behavior of an individual is the result of forces operating simultaneous within his environment in life. Within this theory, attitude toward mathematics is determinants in a learning environment that affect their mathematical performance.

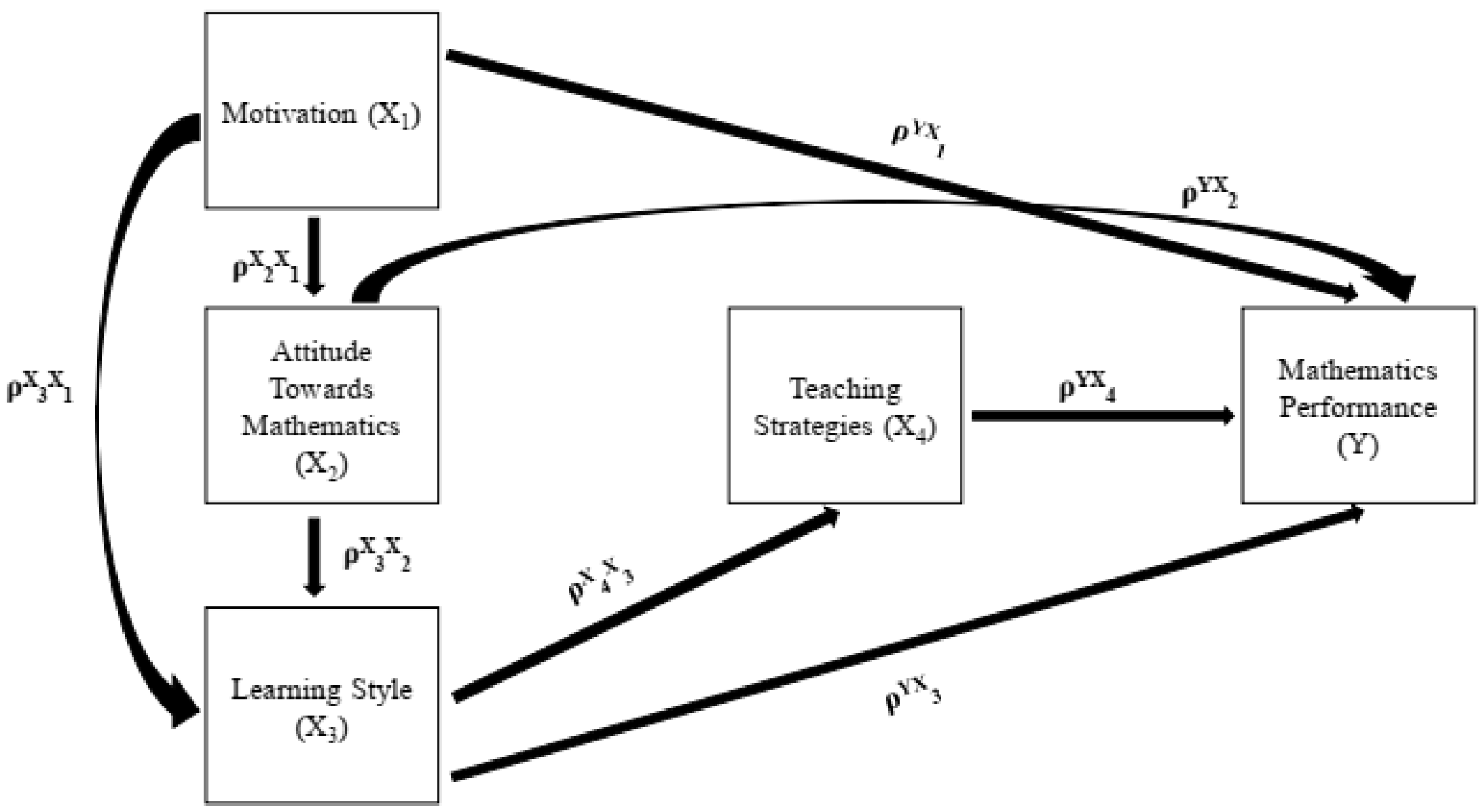

**Figure 1:** Input Path Diagram Representing a Proposed Causal Model

Legend:
$\rho^{YX}{}_1$ – Path coefficient influence of Motivation towards Mathematics Performance
$\rho^{YX}{}_2$ – Path coefficient influence of Attitude towards Mathematics towards Mathematics Performance
$\rho^{YX}{}_3$– Path coefficient influence of Learning Style towards Mathematics Performance
$\rho^{YX}{}_4$– Path coefficient influence of Teaching Strategies towards Mathematics Performance
$\rho^{X_4X_3}$ – Path coefficient influence of Learning Style towards Teaching Strategies
$\rho^{X_3X_1}$ – Path coefficient influence of Motivation towards Learning Style
$\rho^{X_3X_2}$ – Path coefficient influence of Attitude towards Mathematics towards Learning Style
$\rho^{X_2X_1}$ – Path coefficient influence of Motivation towards Attitude towards Mathematics

The causal model in Figure 1 proposed the Mathematics Performance results from Motivation, Attitude towards Mathematics, Learning Style, and Teaching Strategies. Furthermore, the path analysis was carried out through multiple linear regression procedure. The causal model will be predicting the direct and indirect effect.

In this study, the model is specified by the following path equations:

$$Y = \rho^{YX}{}_1X_1 + \rho^{YX}{}_2X_2 + \rho^{YX}{}_3X_3 + \rho^{YX}{}_4X_4 + e_1 \qquad (1)$$

$$X_4 = \rho^{X_3 X_1} X_1 + \rho^{X_4 X_3} X_3 + e_2 \quad (2)$$
$$X_3 = \rho^{X_3 X_1} X_1 + \rho^{X_3 X_2} X_2 + e_3 \quad (3)$$
$$X_2 = = \rho^{X_2 X_1} X_1 + e_2 \quad (4)$$

where:

$Y$ = Mathematics Performace
$X_1$ = Motivation
$X_2$ = Attitude towards Mathematics
$X_3$ = Learning Style
$X_3$ = Teaching Strategies
$\rho^{ij}$ = are the regression coefficient and ;
$e_i$ = is the disturbance term

## METHODOLOGY

Data for this study came from primary and secondary sources. The first was a survey in four instruments namely; motivation (Kusurkar *et al.*, 2011) which focus on the perception level of intrinsic and extrinsic motivation, attitude towards mathematics (Orhun, 2007), learning style (Gilakjani, 2012), and teaching strategies (Hamzeh, 2014). The second part was Mathematics Performance which is the Grade Point Average (GPA) of the respondents.

After collecting the data from all sources, the data were checked for further analysis, that is, checked the restriction of range in the data values, outliers, nonlinearity, and non-normality of data, in order to determine the aptness of the generated model. The path analysis was carried out through the multiple regression procedure in a statistical software. The causal model will be predicting the direct and indirect effect.

## RESULTS

The aptness of the generated model for the said data was evaluated and was tested whether it satisfies the required assumptions. The range of values obtained for variables was considered as a restricted range of one or more variables can reduce the magnitude of relationships. Mahalanobis distance was performed in order to detect the outliers as it can strongly affects the mean and standard deviation of a variable. The linearity assumption was also considered and can best tested with scatter plots, whether the Mathematics Performance are linearly related to the motivation, attitude towards mathematics, learning style and teaching strategies.

Moreover, the analysis requires to check that whether all variables were normally distributed since it determined with a goodness of fit test and affects the resulting Path Analysis. The study used the Kolmogorov-Smirnov test in detecting non-normality. Variance Inflation Factor (VIF) values were also apply to test the multicollinearity assumption, this is to check that the independent variables are not highly correlated with each other. And lastly, the assumption of homoscedasticity were considered through the plot of standardized residuals against predicted values, since it allows to test the variance of error terms in the values of the independent variables. After the basic assumptions had been met, data were analyzed through path analysis.

Figure 2 displays results of the initial path analysis model of Mathematics Performance, motivation, attitude towards mathematics, learning style and teaching strategies. The obtained regression coefficients in Figure 2 was specified by Equation (1 – 4) through multiple regression analysis. Among the four exogenous variables only the motivation factor were not significantly related in the Mathematics Performance (0.047). The motivation were positively related to attitude towards mathematics (0.531*), and to learning style (0.337*). Moreover, learning style were positively related to teaching strategies (0.209*), and to attitude toward mathematics (0.150*).

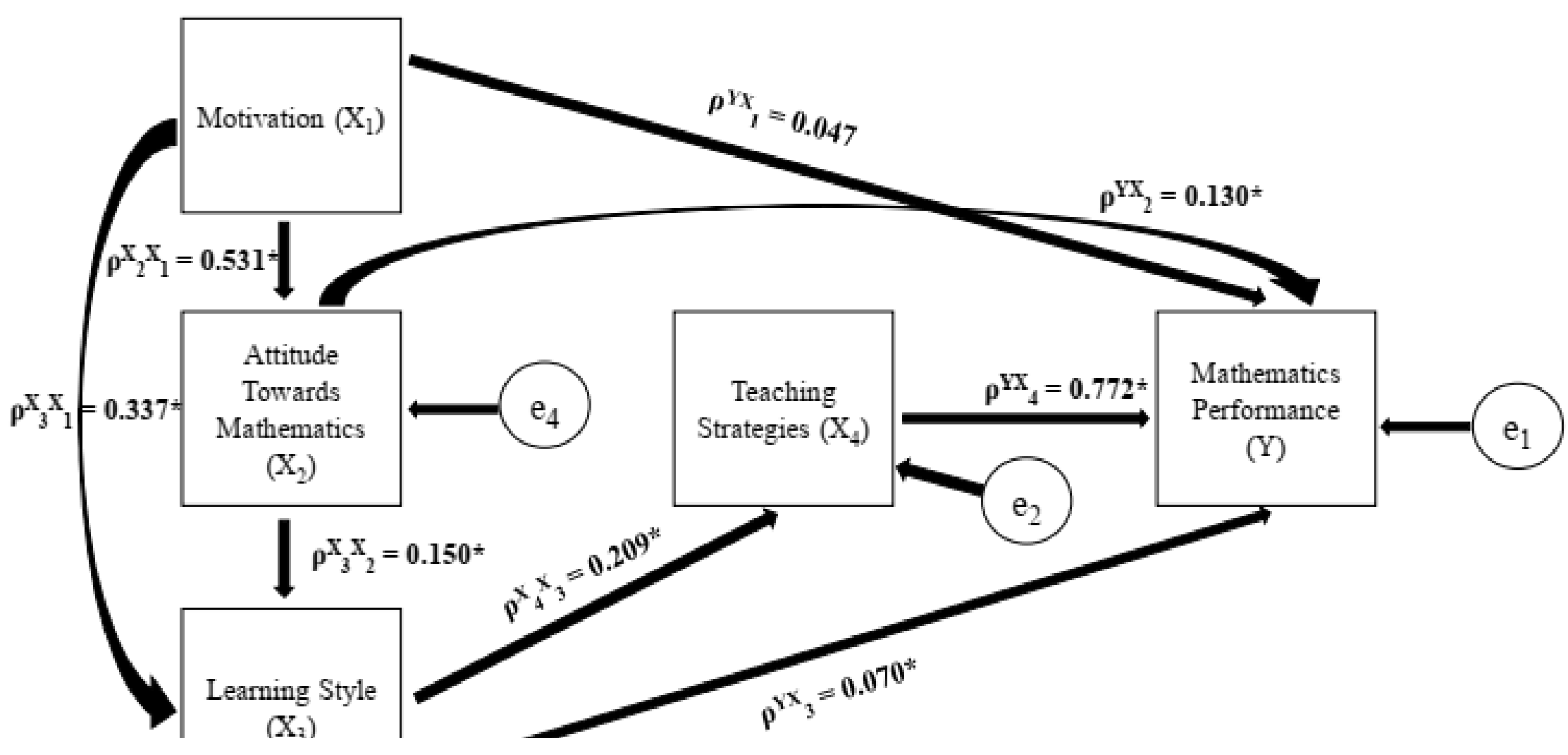

**Figure 2:** Initial Causal Factor Models Affecting the Mathematics Performance

Table 1 shows the calculation of observed correlation for the Mathematics Performance model. The magnitude of the Pearson correlation coefficient determines the strength of the correlation. Although there are no concrete rules for assigning the strength of association to particular values, the study had used the general guideline provided by Cohen (1988):

| Coefficient Value | Strength of Association |
|---|---|
| $0.1 < \lvert r \rvert < 0.3$ | Small Correlation |
| $0.3 < \lvert r \rvert < 0.5$ | Medium/Moderate Correlation |
| $\lvert r \rvert > 0.5$ | Large/Strong Correlation |

Table 1. Calculation of Observed Correlation for the Mathematics Performance Model

| | | $X_1$ | $X_2$ | $X_3$ | $X_4$ | Y |
|---|---|---|---|---|---|---|
| | Pearson Correlation | 1 | .531** | .514** | .466** | .512** |
| $X_1$ | Sig. (2-tailed) | | .000 | .000 | .000 | .000 |
| | N | 240 | 240 | 240 | 240 | 240 |
| | Pearson Correlation | .531** | 1 | .420** | .435** | .520** |
| $X_2$ | Sig. (2-tailed) | .000 | | .000 | .000 | .000 |
| | N | 240 | 240 | 240 | 240 | 240 |
| $X_3$ | Pearson Correlation | .514** | .420** | 1 | .431** | .482** |
| | Sig. (2-tailed) | .000 | .000 | | .000 | .000 |
| | N | 240 | 240 | 240 | 240 | 240 |
| | Pearson Correlation | .466** | .435** | .431** | 1 | .881** |
| $X_4$ | Sig. (2-tailed) | .000 | .000 | .000 | | .000 |
| | N | 240 | 240 | 240 | 240 | 240 |
| | Pearson Correlation | .512** | .520** | .482** | .881** | 1 |
| Y | Sig. (2-tailed) | .000 | .000 | .000 | .000 | |
| | N | 240 | 240 | 240 | 240 | 240 |

**. Correlation is significant at the 0.01 level (2-tailed).

Based on the table, motivation ($X_1$), attitude toward mathematics ($X_2$) and teaching strategies ($X_4$) were positively related to Mathematics Performance (Y), which obtained p-values less than 0.05 level of significance. The strength of association from these variables revealed to have a large/strong correlation. This implies that as the perception level of motivation, attitude towards mathematics and teaching strategies increases, then the Mathematics performance of the students will also increase.

Meanwhile, motivation ($X_1$), attitude towards mathematics ($X_2$), learning style ($X_3$), and teaching strategies ($X_4$) were correlated to each other, which obtained p-values less than 0.05 level of significance. The strength of association from all factors revealed to have a medium/moderate or large/strong correlation.

To evaluate the model fit in Figure 2, obtaining the reproduced correlations and comparing it to the empirical correlations must be needed to assess the consistency of the model.To determine the reproduced correlation between two variables, it involves the identification of all valid paths between the variables in the model. The complete set of path decompositions and reproduced correlations for the model shown in Figure 2 is presented in Table 2. Causal effects are presented by paths consisting only of causal links, that is, only one-headed arrow is used to indicate the effect of a variable pressumed to be the cause on another variable pressumed to be an effect. In this study, a direct effect or a causal path consisting of only one link is denoted by “D”, a direct effect consisting of two or more link is denoted by “I”, and spurious effect that is any path components resulting from paths that have reversed casual direction at some point is denoted by “S”.

Table 2. Calculation of Initial Reproduced Correlation for the Mathematics Performance Model

| Reproduced Correlation | Path Decomposition |
|---|---|
| $\check{r}_{12}$ | $= \rho^{X_2X_1}$ |
| | = **0.531** |
| | (D) |
| $\check{r}_{13}$ | $= \rho^{X_3X_1} + \rho^{X_2X_1}\rho^{X_3X_2}$ |
| | = 0.337 + (0.531)(0.150) = **0.417** |
| | (D) (I) |
| $\check{r}_{14}$ | $= \rho^{X_2X_1}\rho^{X_3X_2}\rho^{X_4X_3} + \rho^{X_3X_1}\rho^{X_4X_3}$ |
| | = (0.531)(0.150)(0.209) + (0.337)(0.209) = **0.087** |
| | (I) (I) |
| $\check{r}_{15}$ | $= \rho^{YX_1} + \rho^{X_2X_1}\rho^{YX_2} + \rho^{X_2X_1}\rho^{X_3X_2}\rho^{YX_3} + \rho^{X_2X_1}\rho^{X_3X_2}\rho^{X_4X_3}\rho^{YX_4} + \rho^{X_3X_1}\rho^{YX_3} + \rho^{X_3X_1}\rho^{X_4X_3}\rho^{YX_4}$ |
| | = 0.047 + (0.531)(0.130) + (0.531)(0.150)(0.070) + (0.531)(0.150)(0.209)(0.772) + |
| | (D) (I) (I) (I) |
| | (0.337)(0.070) + (0.337)(0.209)(0.722) = **0.261** |
| | (I) (I) |
| $\check{r}_{23}$ | $= \rho^{X_3X_2} + \rho^{X_2X_1}\rho^{X_3X_1}$ |
| | = 0.150 + (0.531)(0.337) = **0.329** |
| | (D) (S) |
| $\check{r}_{24}$ | $= \rho^{X_3X_2}\rho^{X_4X_3} + \rho^{X_2X_1}\rho^{X_3X_1}\rho^{X_4X_3}$ |
| | = (0.150)(0.209) + (0.531)(0.337)(0.209) = **0.069** |
| | (I) (S) |
| $\check{r}_{25}$ | $= \rho^{YX_2} + \rho^{X_3X_2}\rho^{YX_3} + \rho^{X_3X_2}\rho^{X_4X_3}\rho^{YX_4} + \rho^{X_2X_1}\rho^{X_3X_1}\rho^{YX_3} + \rho^{X_2X_1}\rho^{X_3X_1}\rho^{X_4X_3}\rho^{YX_4}$ |
| | = 0.130 + (0.150)(0.070) + (0.150)(0.209)(0.772) + |
| | (0.531)(0.337)(0.070) +<br>(D) (I) (I) (S)<br>(0.531)(0.337)(0.209)(0.772) = **0.206**<br>(S) |
| *ř*$\check{r}_{34}$ | $= \rho^{X}{}_{4}{}^{X}{}_{3}$<br>= **0.209**<br>(D) |
| $\check{r}_{35}$ | $=\rho^{YX}{}_{3} + \rho^{X}{}_{4}{}^{X}{}_{3}\,\rho^{YX}{}_{4} + \rho^{X}{}_{3}{}^{X}{}_{1}\,\rho^{YX}{}_{1} + \rho^{X}{}_{3}{}^{X}{}_{1}\,\rho^{X}{}_{2}{}^{X}{}_{1}\,\rho^{YX}{}_{2} + \rho^{X}{}_{3}{}^{X}{}_{2}\,\rho^{YX}{}_{2}$<br>= 0.070 + (0.209)(0.772) + (0.337)(0.047) + (0.337)(0.531)(0.130) + (0.150)(0.130)<br>(D) (I) (S) (S) (S)<br>=**0.290** |
| $r_{45}$ | $= \rho^{YX}{}_{4} + \rho^{X}{}_{4}{}^{X}{}_{3}\,\rho^{YX}{}_{3}$<br>= 0.772 + (0.209)(0.070) = **0.787**<br>(D) (S) |

D - direct effect, I – indirect effect, S – spurious effect

To obtain the reproduced correlation (initial model) in Table 3, the set of legitimate paths in Table 2 was used, that is, making the substitutions of path coefficients in Figure 2. In assessing the fit of the model in Figure 2, it can be gleaned from Table 3 that seven out of ten reproduced correlations have differences greater than 0.05. Hence, those reproduced correlations that have difference greater than 0.05 from the empirical correlations indicate that the model is not consistent with the empirical data, thus, it was determined that several paths should be added in the model and revisions to the model are warranted prior to describing any of the causal effects, since, there are one or more missing paths in the model. Moreover, because the beta coefficient of the path from $X_1$ to $X_5$ is not significant, then it was removed to form the revised model (final).

Table 3. Observed and Initial Reproduced Correlations for the Mathematics Performance Model

| | $X_1$ | $X_2$ | $X_3$ | $X_4$ | Y |
|---|---|---|---|---|---|
| | | | Observed Correlation | | |
| $X_1$ | 1 | | | | |
| $X_2$ | 0.531 | 1 | | | |
| $X_3$ | 0.514 | 0.420 | 1 | | |
| $X_4$ | 0.466 | 0.435 | 0.431 | 1 | |
| Y | 0.512 | 0.520 | 0.482 | 0.881 | 1 |
| | | | Reproduced Correlation | | |
| $X_1$ | 1 | | | | |
| $X_2$ | 0.531 | 1 | | | |
| $X_3$ | 0.417 | 0.329* | 1 | | |
| $X_4$ | 0.087* | 0.069* | 0.209* | 1 | |
| Y | 0.261* | 0.206* | 0.482 | 0.787* | 1 |

*Difference between reproduced and observed correlation is greater than 0.05.

The revised path diagram, including path coefficients is presented in Figure 3,

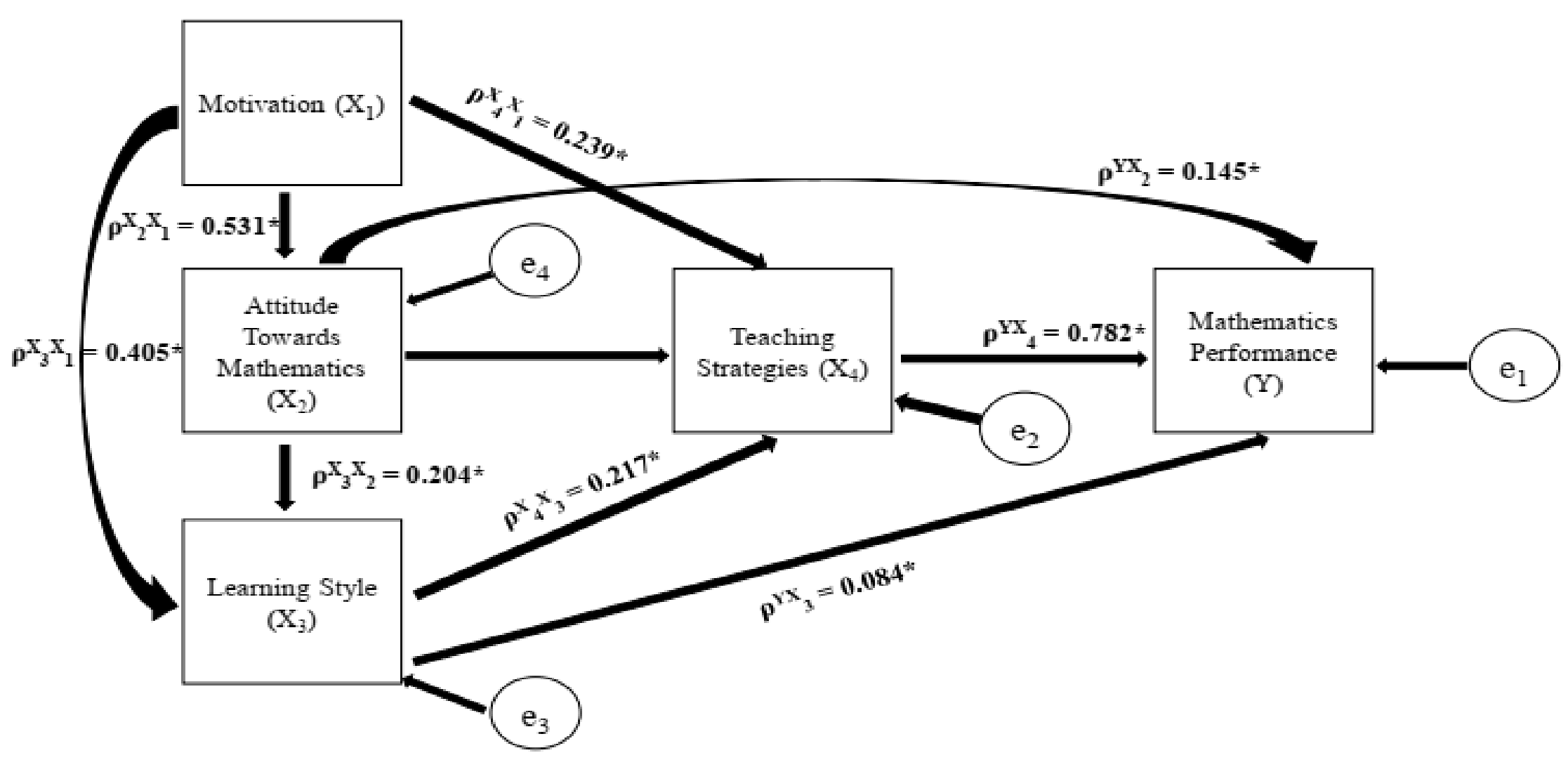

**Figure 3**. Revised Causal Factor Models Affecting the Mathematics Performance

Once a model has been revised, the fit should again be reassessed in order to generate the best model. The results of obtaining the complete set of paths for the revised model is given by Table 4. It can be seen from table 5 that the results of revised model is much better fit than the initial model that is the there is only one reproduced correlation that have difference greater than 0.05 from observed correlation**.**

Table 4. Calculation of Revised Reproduced Correlation for the Mathematics Performance Model

| Reproduced Correlation | Path Decomposition |
|---|---|
| $\check{r}_{12}$ | $= \rho_{X_2X_1}$<br>= **0.531**<br>(D) |
| $\check{r}_{13}$ | $= \rho_{X_3X_1} + \rho_{X_2X_1}\rho_{X_3X_2}$<br>= *0.405* + (0.531)0.204) = **0.513**<br>(D)(I) |
| $\check{r}_{14}$ | $= \rho_{X_4X_1} + \rho_{X_2X_1}\rho_{X_3X_2}\rho_{X_4X_3} + \rho_{X_2X_1}\rho_{X_4X_2}\rho_{X_3X_1}\rho_{X_4X_3}$<br>= 0.239 + (0.531)(0.204)(0.217) + (0.531)(0.216)(0.405)(0.217) = **0.465**<br>(D) (I) (I) |
| $\check{r}_{15}$ | $= \rho_{X_2X_1}\rho_{YX_2} + \rho_{X_2X_1}\rho_{X_3X_2}\rho_{YX_3} + \rho_{X_2X_1}\rho_{X_3X_2}\rho_{X_4X_3}\rho_{YX_4} + = \rho_{X_3X_1}\rho_{YX_3} + \rho_{X_3X_1}\rho_{X_4X_3}\rho_{YX_4} + \rho_{X_4X_2}\rho_{YX_4}$<br>= (0.531)(0.145) + (0.531)(0.204)(0.084) + (0.531)(0.204)(0.217)(0.782) +<br>(I) (I) (I)<br>(0.405)(0.084) + (0.405)(0.217)(0.782) + (0.239)(0.782) = **0.394**<br>(I) (I) (I) |
| $\check{r}_{23}$ | $= \rho_{X_3X_2} + \rho_{X_2X_1}\rho_{X_3X_1}$<br>= 0.204 + (0.531)(0.405) = **0.419**<br>(D) (S) |
| $\check{r}_{24}$ | $= \rho_{X_4X_2} + \rho_{X_3X_2}\rho_{X_4X_3} + \rho_{X_2X_1}\rho_{X_4X_1} + \rho_{X_2X_1}\rho_{X_3X_1}\rho_{X_4X_3}$ |

= 0.216 + (0.204)(0.217) + (0.531)(0.239) + (0.531)(0.405)(0.217) = **0.434**

(D)(I) (S) (S)

*ř25* $= \rho^{YX_2} + \rho^{X_3X_2}\rho^{YX_3} + \rho^{X_3X_2}\rho^{X_4X_3}\rho^{YX_4} + \rho^{X_4X_2}\rho^{YX_4} + \rho^{X_2X_1}\rho^{X_3X_1}\rho^{YX_3} + \rho^{X_2X_1}\rho^{X_3X_1}\rho^{X_4X_3}\rho^{YX_4} + \rho^{X_2X_1}\rho^{X_4X_1}\rho^{YX_4}$

= 0.145 + (0.204)(0.084) + (0.204)(0.217)(0.782) + (0.216)(0.782) +

(D) (I) (I) (I)

(0.531)(0.405)(0.084) + (0.531)(0.405)(0.217)(0.782) + (0.531)(0.239)(0.782) = **0.519**

(S) (S) (S)

---

*ř34* $= \rho^{X_4X_3} + \rho^{X_3X_1}\rho^{X_4X_1} + \rho^{X_3X_1}\rho^{X_2X_1}\rho^{X_4X_2} + \rho^{X_3X_2}\rho^{X_4X_2}$

= 0.217 + (0.405)(0.239) + (0.405)(0.531)(0.216) + (0.204)(0.216) = **0.404**

(D) (S) (S) (S)

*ř35* $= \rho^{YX_3} + \rho^{X_4X_3}\rho^{YX_4} + \rho^{X_3X_1}\rho^{X_2X_1}\rho^{YX_2} + \rho^{X_3X_1}\rho^{X_2X_1}\rho^{X_4X_2}\rho^{YX_4} + \rho^{X_3X_1}\rho^{X_4X_2}\rho^{YX_4} +$

$\rho^{X_3X_2}\rho^{YX_2} + \rho^{X_3X_2}\rho^{X_4X_2}\rho^{YX_4}$

= 0.084 + (0.217)(0.782) + (0.405)(0.531)(0.145) + (0.405)(0.531)(0.216)(0.782) +

(D) (I) (S) (S)

(0.405)(0.239)(0.782) + (0.204)(0.145) + (0.204)(0.216)(0.782) = **0.468**

(S) (S) (S)

---

*ř45* $= \rho^{YX_4} + \rho^{X_4X_1}\rho^{X_2X_1}\rho^{YX_2} + \rho^{X_4X_1}\rho^{X_2X_1}\rho^{X_3X_2}\rho^{YX_3} + \rho^{X_4X_1}\rho^{X_2X_1}\rho^{X_3X_2}\rho^{X_4X_3}\rho^{YX_4} +$

$\rho^{X_4X_1}\rho^{X_2X_1}\rho^{X_4X_2}\rho^{YX_4}$

= 0.782 + (0.239)(0.531)(0.145) + (0.239)(0.531)(0.204)(0.084) +

(D) (S) (S)

(0.239)(0.531)(0.204)(0.217)(0.782) +

---

(0.239)(0.531)(0.216)(0.782) +

(S) (S)

(0.239)(0.405)(0.084) + (0.239)(0.405)(0.217)(0.782) + (0.216)(0.145) +

(S) (S) (S)

(0.216)(0.204)(0.084) + (0.217)(0.084) = **0.906**

(S) (S)

D - direct effect, I – indirect effect, S – spurious effect

Table 5. Observed and Revised Reproduced Correlations for the Mathematics Performance Model

| | $X_1$ | $X_2$ | $X_3$ | $X_4$ | Y |
|---|---|---|---|---|---|
| | | | Observed Correlation | | |
| $X_1$ | 1 | | | | |
| $X_2$ | 0.531 | 1 | | | |
| $X_3$ | 0.514 | 0.420 | 1 | | |
| $X_4$ | 0.466 | 0.435 | 0.431 | 1 | |
| Y | 0.512 | 0.520 | 0.482 | 0.881 | 1 |
| | | | Reproduced Correlation | | |
| $X_1$ | 1 | | | | |
| $X_2$ | 0.531 | 1 | | | |
| $X_3$ | 0.513 | 0.419 | 1 | | |
| $X_4$ | 0.465 | 0.434 | 0.404 | 1 | |
| Y | 0.394* | 0.519 | 0.468 | 0.906 | 1 |

*Difference between reproduced and observed correlation is greater than 0.05.

Mathematics had indirect effects on Motivation (0.394), Attitude toward Mathematics (0.221), and Learning Style (0.084). Teaching Strategies had indirect effect on Motivation (0.226), and Attitude towards Mathematics (0.044). Finally, learning style had indirect effect on Motivation (0.108).

R-squared indicated that 80.5% of the variance in the Mathematics Performance was accounted for the independent variables. The model explains 29.9% of the variance in teaching strategies, 29.4% of the variance in learning style and, finally, 28.2% in attitudes towards mathematics.

Table 6. Summary of Causal Effects for Revised Model (Mathematics Performance)

| Outcome | Determinants | Causal Effects | | |
|---|---|---|---|---|
| | | Direct | Indirect | Total |
| Mathematics Performance ($R^2 = 0.805$) | Motivation | - | 0.394 | 0.394 |
| | Attitude Towards Mathematics | 0.145 | 0.221 | 0.366 |
| | Learning Style | 0.170 | 0.084 | 0.254 |
| | Teaching Strategies | 0.782 | - | 0.782 |
| Teaching Strategies ($R^2 = 0.299$) | Motivation | 0.239 | 0.226 | 0.465 |
| | Attitude Towards Mathematics | 0.216 | 0.044 | 0.260 |
| | Learning Style | 0.217 | - | 0.217 |
| Learning Style ($R^2 = 0.294$) | Motivation | 0.405 | 0.108 | 0.513 |
| | Attitude Towards Mathematics | 0.204 | - | 0.204 |
| Attitude Towards Mathematics | Motivation | 0.531 | - | 0.531 |

($R^2 = 0.282$)

## DISCUSSION

The main purpose of the study was concerned in direct and indirect effect of student's motivation, attitude towards mathematics, learning style, and perception of the students teaching strategies of the teachers to Mathematics Performance.

Generally, Mathematics Performance had a positive direct effect on attitudes towards mathematics, learning style, and a very strong relation was found between mathematics performance and teaching strategies. Furthermore, motivation, attitude towards mathematics and learning style affected learning style positively. Finally, learning style was directly related to motivation and attitude towards mathematics, and motivation directly positively related to attitude towards mathematics. The study of Areepattamannil and Kaur (2012) found out that positive affect towards mathematics influenced in mathematics achievement and, it has a great impact to students' mathematics achievement when teaching strategy coincide with the students' learning style (Tebabal & Kahssay, 2011).

## CONCLUSIONS AND RECOMMENDATIONS

The present study provided an empirical test of a causal model concerning relationships among motivation, attitude towards mathematics, learning style, and teaching strategies, and the mathematics performance. As relationships, the strength of association from all factors of mathematics performance revealed to have a medium/moderate or large/strong correlation. Results also reveal that a large proportion of mathematics performance can be directly predicted from attitude towards mathematics, learning style, and teaching strategies. Moreover, attitude towards mathematics, learning style, and teaching strategies influence mathematics performance in direct and indirect ways.

In conclusion, if the students build their attitude toward mathematics, they would be more likely to do their learning style and cope up with the teaching strategies of the teachers and therefore enhanced their mathematics performance.

## FURTHER STUDY

This research concerned only on the factors affecting the mathematics performance of the students such as; student's motivation, attitude towards mathematics, learning style, and perception of the students teaching strategies of the teachers. Since it is widely known that are so many factors on enhancing the mathematics performance of the students, hence, the researchers wish the future researchers to include the factors not considered in this study so that the results will be more in-depth and more complex.

## ACKNOWLEDGMENT

The researchers of this study would like to thank the Agusan del Sur State College of Agriculture and Technology for allowing us to conduct and complete the study.